\documentclass[english,prl,twocolumn,showpacs,amsmath,amssymb,superscriptaddress,floatfix]{revtex4}

\usepackage[T1]{fontenc}
\usepackage{graphicx}
\usepackage[latin9]{inputenc}
\usepackage{amsmath}
\usepackage{amssymb}
\usepackage{esint}

\makeatletter
\gdef\@ptsize{0}
\let\@cursize\normalize
\makeatother
\usepackage{setspace}

\usepackage{bm}
\usepackage{pifont}
\usepackage{babel}

\usepackage{wrapfig}

\makeatother

\begin{document}

\preprint{}

\title{Quantum paraelectric varactors for radio-frequency measurements at mK temperatures}

\author{P. Apostolidis}
\affiliation{London Centre for Nanotechnology, University College London, London WC1H 0AH, United Kingdom}
\affiliation{Department of Physics and Astronomy, University College London, London WC1E 6BT, United Kingdom}

\author{B. J. Villis}
\affiliation{London Centre for Nanotechnology, University College London, London WC1H 0AH, United Kingdom}

\author{J. F. Chittock-Wood}
\affiliation{Department of Physics and Astronomy, University College London, London WC1E 6BT, United Kingdom}

\author{A. Baumgartner}
\affiliation{Department of Physics, University of Basel, Klingelbergstrasse 82, 4056 Basel, Switzerland}

\author{V. Vesterinen}
\affiliation{VTT Technical Research Centre of Finland, P.O. Box 1000, FI-02044 VTT, Finland}

\author{S. Simbierowicz}
\affiliation{VTT Technical Research Centre of Finland, P.O. Box 1000, FI-02044 VTT, Finland}

\author{J. Hassel}
\affiliation{VTT Technical Research Centre of Finland, P.O. Box 1000, FI-02044 VTT, Finland}

\author{M. R. Buitelaar}
\email{m.buitelaar@ucl.ac.uk}
\affiliation{London Centre for Nanotechnology, University College London, London WC1H 0AH, United Kingdom}
\affiliation{Department of Physics and Astronomy, University College London, London WC1E 6BT, United Kingdom}

\date{\today}

\begin{abstract}
Radio-frequency reflectometry allows for fast and sensitive electrical readout of charge and spin qubits hosted in quantum dot devices coupled to resonant circuits. Optimizing readout, however, requires frequency tuning of the resonators and impedance matching. This is difficult to achieve using conventional semiconductor or ferroelectric-based varactors in the detection circuit as their performance degrades significantly in the mK temperature range relevant for solid-state quantum devices. Here we explore a different type of material, strontium titanate, a quantum paraelectric with exceptionally large field-tunable permittivity at low temperatures. Using strontium titanate varactors we demonstrate perfect impedance matching and resonator frequency tuning at 6 mK and characterize the varactors at this temperature in terms of their capacitance tunability, dissipative losses and magnetic field sensitivity. We show that this allows us to optimize the radio-frequency readout signal-to-noise ratio of carbon nanotube quantum dot devices to achieve a charge sensitivity of 4.8 $\mu$e/Hz$^{1/2}$ and capacitance sensitivity of 0.04 aF/Hz$^{1/2}$.
\end{abstract}


\maketitle
% 73.63.Kv Quantum dots
% 73.63.Fg Nanotubes
% 73.23.Hk Coulomb blockade; single-electron tunneling
% 73.21.La Quantum dots
% 03.67.Lx Quantum computation architectures and implementations

% ------------------------------------------------------------------
% Main text
% -------------------------------------------------------------------
%

Radio-frequency (rf) reflectometry is a measurement technique that allows fast and sensitive readout of charge detectors such as single-electron transistors \cite{Schoelkopf1998,Aassime2001,Brenning2006} and quantum point contacts \cite{Cassidy2007,Reilly2007}, as well as quantum dot devices that host charge or spin qubits \cite{Petersson2010,Chorley2012,Jung2012,Colless2013,GonzalezZalba2015,Harabula2017,Ahmed2018}. By coupling the quantum devices to lumped-element electrical resonators, or on-chip stub tuners \cite{Hasler2015}, it is possible, in principle, to perfectly impedance match the devices to the rf feedlines that connect to them \cite{Pozar2012}. This ensures an optimum power transfer to the devices and the best charge and capacitance sensitivities are expected in this readout regime \cite{Muller2013}. However, since there is often large and unpredictable variability between device impedances (including parasitics), it is necessary to have in-situ tunability of the resonator circuit to ensure perfect matching. Moreover, to achieve optimal performance when using rf components in the detection circuit, such as low-noise amplifiers, which have a narrow operation bandwidth \cite{Simbierowicz2018} or for multiplexing signals of several readout channels \cite{Hornibrook2014}, tunability of the resonant frequency of the circuit is required.

In situ tunability for achieving impedance matching and frequency tuning is feasible with voltage-tunable capacitors, also known as varactors. However, commercially available varactors are not designed for low-temperature operation. While semiconductor-based varactors have been demonstrated to work down to temperatures of around 1 K \cite{Muller2010,Hellmuller2012,House2016,Ares2016,Ibberson2018}, their performance degrades significantly in the mK temperature range - at which most solid-state quantum devices are operated - due to the freezing out of free charge carriers.

Another conventional type of varactor is based on ferroelectric materials which have a high dielectric permittivity that can be tuned by an electric field. Examples of ferroelectrics commonly found in varactors are those in the lead titanate and barium strontium titanate families of solid solutions. Barium strontium titanate (Ba$_{1-x}$Sr$_x$TiO$_3$ or BST) is a particularly well-studied material which has good tunabilty at room temperature and relatively low loss tangent. These materials, however, are typically operated in their paraelectric state - that is, well above their Curie temperature - to minimize losses. At low temperatures, materials such as BST become ferroelectric, lose their tunability, and dissipative losses increase.

\begin{figure*}
\includegraphics[width=178mm]{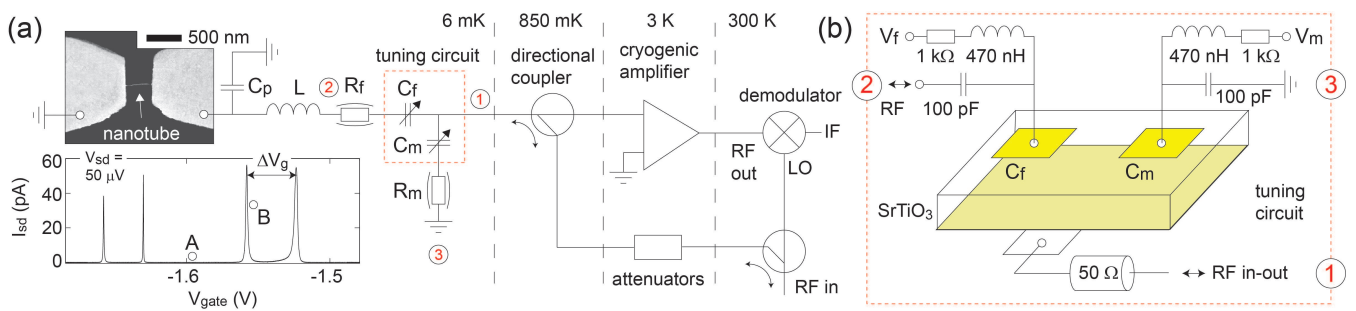}
\caption{\label{Fig1} \textbf{(a)} Simplified schematic of the radio-frequency detection circuit. The scanning electron micrograph shows a carbon nanotube quantum dot device which is embedded in an $LC$ resonant circuit for high sensitivity and high bandwidth measurements. Two SrTiO$_3$ varactors are incorporated in the circuit for frequency tuning and impedance matching. The rf signals are amplified using a low-noise cryogenic amplifier operating at 3 K. Room temperature demodulation using a vector network analyzer provides a measurement of both quadratures. The inset shows dc conductance measurements of the nanotube device (using a bias tee not shown in the schematic). \textbf{(b)} The two varactors are fabricated on a single SrTiO$_3$ crystal and have a parallel plate geometry. The varactors are voltage biased using bias-tees and characterized in the schematic of panel (a) by a variable capacitance and effective series resistance to incorporate losses.}
\end{figure*}

In this work we therefore take a different approach and explore quantum paraelectricity, a phenomenon in which ferroelectric order is suppressed by quantum fluctuations down to zero kelvin, as the basis for a mK varactor. Only a few materials, such as SrTiO$_3$, KTaO$_3$, and CaTiO$_3$, are known to be quantum paraelectrics. Of these, SrTiO$_3$ in particular has an extremely high relative permittivity, of order 10,000, at temperatures below $\sim$ 4K \cite{Saifi1970,Sakudo1971,Neville1972,Rowley2014,Davidovikj2017}. Important for our application, the relative permittivity of SrTiO$_3$ is tunable using electric fields by over an order of magnitude in this temperature range. We take advantage of these properties and show that SrTiO$_3$ varactors are easily fabricated and, when integrated in an rf measurement circuit, allow frequency tuning and perfect impedance matching down to 6 mK, the lowest achievable temperature in our measurement system. We characterize the performance of the varactors at this temperature in terms of their capacitance tunability, dissipative losses and magnetic field dependence. To demonstrate their use for the rf readout of quantum devices we measure a carbon nanotube quantum dot charge detector, and show that the SrTiO$_3$ varactors allow us to optimize the signal-to-noise ratio to achieve a charge sensitivity of 4.8 $\mu$e/Hz$^{1/2}$.

\section{Results}

\textbf{Matching network and device} The circuit we consider is shown schematically in Fig.~1a. It consists of a carbon nanotube quantum dot coupled to an inductor $L=330$ nH and parasitic capacitance $C_p = 3.2$ pF. Two SrTiO$_3$ varactors are included for frequency tuning and impedance matching as illustrated in Fig.~1b. The carbon nanotube is metallic (narrow band-gap) and grown on a degenerately doped Si substrate terminated by 280 nm of SiO$_2$. The capacitive coupling between the nanotube contact pads and Si substrate dominates the  parasitic capacitance of the device. The carbon nanotube device has a length of 400 nm between its source and drain contacts and a room-temperature resistance of 280 k$\Omega$. At mK temperatures, the nanotube behaves as a quantum dot and transport is dominated by Coulomb blockade and level quantization. Using bias-tees (not shown in the schematic) we are able to measure the dc transport characteristic of the nanotube quantum dot, see graph in Fig.~1a, using a source-drain bias voltage $V_{sd}= 50$ $\mu$V. We observe a pronounced pairing of conductance peaks indicative of a two-fold spin degenerate electron system. Fourfold grouping of conductance peaks is not observed for this nanotube device \cite{Laird2015}. From the dc measurements we determine the charging energy $U_C \sim 7$ meV and single-particle level spacing $\Delta E \sim 4$ meV.

The varactors consist of parallel metallic pads (5/60 nm Ti/Au) on either side of a 0.5 mm TiO$_2$-terminated single-crystal SrTiO$_3$ (001) substrate. The top pads are 120x120 $\mu$m squares that are voltage biased against the fully metallised bottom of the substrate. The latter is directly connected to the rf measurement line. Both varactors are thus fabricated on the same SrTiO$_3$ substrate which has a total area of 3x3 mm$^2$ and are separated by about 2 mm.

The circuit is measured in a dilution refrigerator with a base (lattice) temperature of 6 mK. All dc lines are filtered at various stages with a filter bandwidth of $\sim 3$ kHz. This yields an electron temperature of 12 mK which is measured using the carbon nanotube quantum dot as a primary thermometer. We use dc control voltages $V_f$ and $V_m$ applied via bias-tees to tune the varactors. The rf signal is attenuated at various stages along the dilution refrigerator and applied to the device and matching circuit using a directional coupler. The reflected signal is amplified using a low-noise amplifier with noise temperature $T_N \sim 5$ K mounted at the 3 K stage and demodulated at room temperature using a vector network analyzer.

\begin{figure*}
\includegraphics[width=170mm]{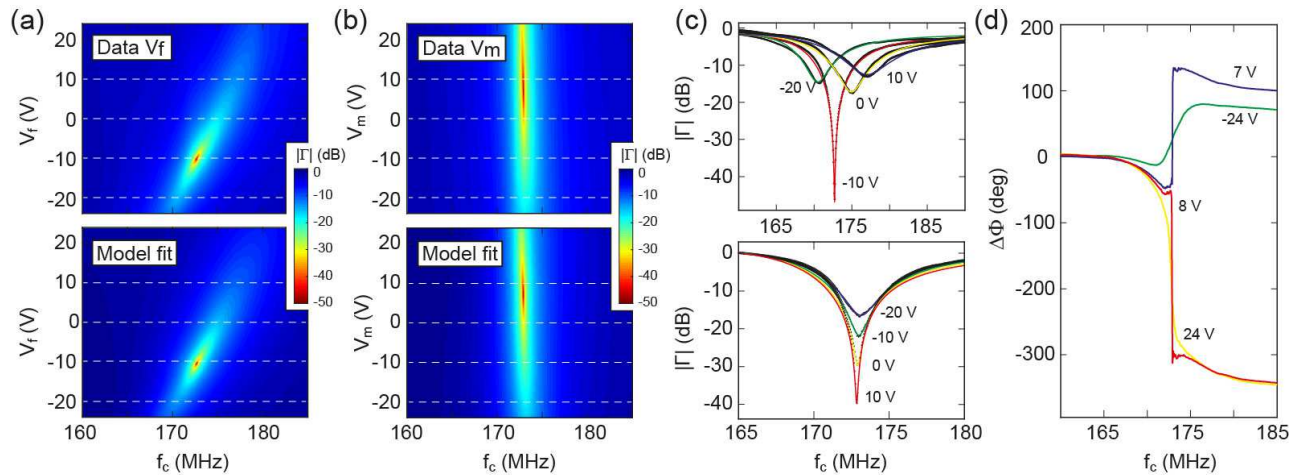}
\caption{\label{Fig2} \textbf{(a)} Colourscale plot of the measured reflection coefficient magnitude $|\Gamma|$ (top) and model fits (bottom) as a function of rf frequency and varactor voltage $V_f$. For these measurements $V_m$ is set to zero. The data shows a shift of the resonance frequency to higher values for more positive $V_f$. The simultaneous change in matching conditions results in a strong variation of the measured magnitude. Perfect matching is observed around $V_f \sim -10$ V. \textbf{(b)} Colourscale plot of the measured reflection coefficient magnitude $|\Gamma|$ (top) and model fits (bottom) as a function of rf frequency and varactor voltage $V_m$. For these measurements, $V_f$ is set to zero. Changing $V_m$ strongly affects the magnitude response but not the resonance frequency. \textbf{(c)} Measured data and model fits of the reflection coefficient magnitude $|\Gamma|$ along the dashed lines indicated in panels (a) and (b). \textbf{(d)} Measured rf phase response along the dashed lines indicated in panel (b). The data show a clear transition from under to overcoupling, with perfect matching observed around $V_m=7.5$ V.}
\end{figure*}

\textbf{Impedance matching and frequency tuning} We first characterise the SrTiO$_3$ varactors by tuning the carbon nanotube quantum dot in a Coulomb blockade regime where the conductance is absent (position 'A' in the graph of Fig.~1a). Figures 2a and 2b show the resulting rf amplitude response as a function of frequency and control voltages $V_f$ and $V_m$, respectively. In Fig. 2a we set $V_m=0$ V and vary $V_f$ between -24 and 24 V. This results in a pronounced shift of the resonance frequency over $\sim 12$ MHz. At the same time, the strength of the amplitude response changes, as also seen in the line scans of Fig.~2c (upper panel), reflecting a simultaneous change in the matching conditions, with perfect matching observed around $V_f = -10$ V. In Fig. 2b we set $V_f=0$ V and vary $V_m$ between -24 and 24 V. In this case, there is little change in the resonance frequency but a pronounced change in the matching conditions. Perfect matching is observed around $V_m = 7.5$ V, at which point the system changes from over to under coupled to the rf feedline, most clearly seen in the phase data of Fig.~2d.

To quantitatively understand the response of the SrTiO$_3$ varactors we parameterize the varactors by their voltage-dependent capacitance and effective series resistance which takes into account dissipative losses. Using these two variables as fit parameters, we are able to obtain excellent agreement with the observed data, see the bottom panels of Figs.~2a and 2b, and Supplementary Section I for further details on the fit procedure. The results are summarized in Fig.~3a which shows the values obtained for $C_f$ and $R_f$ as a function of the control voltage $V_f$. We observe a monotonically decreasing capacitance with a tunability from about 40 pF to 15 pF over our voltage range. At the same time we observe a monotonically increasing effective series resistance changing from about 2.5 $\Omega$ to 16 $\Omega$ with the notion that these values also include contributions from the chip inductor and circuit board so should be considered an upper bound for the losses in the varactor.

These measurements are consistent with the behaviour of SrTiO$_3$ at higher temperatures and show that the material retains its characteristic tunability down to mK temperatures. The voltage at which the maximum capacitance is observed is outside of the range of the measurements in Fig.~2. By increasing the control voltage to larger negative values, see Fig.~3b, a maximum capacitance - resulting in the lowest observed resonant frequency $f_0$ - is seen at $V_f \sim -45$ V. The maximum capacitance at this point is approximately 52 pF which, using finite-element simulations of our device geometry, implies a dielectric constant $\epsilon^{STO}_{r,max}= 23,000$, in good agreement with expectations for SrTiO$_3$ \cite{Vendik1998}.

The extracted values for the capacitance and effective series resistance also allow us to estimate the dielectric loss tangent of the varactor defined as tan$(\delta) = \epsilon^{''}/\epsilon^{'}$, where $\epsilon^{''}$ and $\epsilon^{'}$ are the imaginary and real part of the relative permittivity \cite{Pozar2012}. The loss tangent can be related to the effective series resistance tan$(\delta) = R_f C_f \omega$, where $\omega =2 \pi f$ is the angular frequency of the measurement. Since in our experiments we can not distinguish between loss contributions from the SrTiO$_3$ varactors and the chip inductor and circuit board, it is difficult to provide an exact value for the varactor loss tangent. Nevertheless, from the values for $R_f$ and $C_f$ extracted in Fig.~3a, we can estimate tan$(\delta)$ to be in the $10^{-1} - 10^{-2}$ range for large negative voltages where $C_f$ has a maximum. Dissipation in SrTiO$_3$ has been attributed to the interaction between the oscillating (rf) electric field and acoustic phonons, where the presence of a large dc electric field breaks the crystal lattice symmetry and introduces a field-dependent dipole moment in the unit cell, making the system more susceptible to rf losses \cite{Ferroelectrics2009}. In single crystals, this results in an increase of the observed loss tangent for larger dc electric fields which is consistent with our data.

The shift of the maximum capacitance to negative voltages as seen in Fig,~3b has previously been reported (see e.g. Ref. \cite{Caviglia2008}) and has been attributed to trapped charges such as oxygen vacancies in the crystal. This also gives rise to hysteretic behaviour and a dependence of the capacitance on the voltage sweep history. This is seen, for example, in the line traces of $V_f$ and $V_m$ taken at 0 V in Fig.~2a,b which show a slightly different response even though taken under nominally identical conditions. This does not affect device stability which we verified by measuring the phase and magnitude response at the rf resonance frequency over time. More details on the observed hysteresis and varactor stability at mK temperatures are provided in Supplementary Section IV.

For operation in a reflectometry setup, it is important to be able to tune both the resonant frequency and impedance matching independently. This is demonstrated in Fig.~3c where we operate both SrTiO$_3$ varactors. We first use $V_f$ to set the resonant frequency followed by applying a suitable voltage on $V_m$ to obtain perfect matching. Using the two varactors we are able to set both the resonant frequencies and obtain impedance matching in a frequency window between 167 MHz and 182 MHz for control voltages $V_{f,m} < 50$ V. The obtained frequency shift is larger than the bandwidth of the resonances, which is relevant in applications for which multiplexing is important. Additionally, we tested the magnetic field dependence of the SrTiO$_3$ varactors as shown in Fig.~3d for the phase response. No change is observed over the measurement range up to 9 T for both the resonance frequency and matching. This is important for applications in which a magnetic field is used to tune device characteristics, such as spin-based quantum information processing. Using SrTiO$_3$ varactors, any variation in the measured amplitude or phase can then confidently be attributed to the tested devices rather than changes in the matching or detection circuit.

\begin{figure}
\includegraphics[width=87mm]{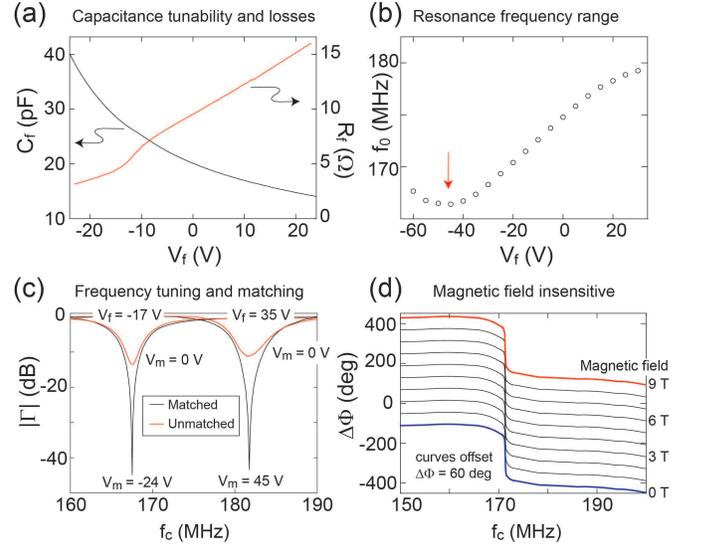}
\caption{\label{Fig3} \textbf{(a)} Varactor capacitance $C_f$ and effective series resistance $R_f$ as a function of tuning voltage $V_f$ as deduced from the model fits in Fig.~2. The varactor shows a monotonic decrease (increase) of $C_f$ ($R_f$) with increasing positive $V_f$ from -24 V to 24 V. \textbf{(b)} Measured over a larger voltage range, the varactor yields a minimum resonance frequency (corresponding to a maximum $C_f$) for large negative voltage. Here this minimum is observed around $V_f = - 45$ V, see arrow, after which the resonance frequency increases again. \textbf{(c)} The resonance frequency and impedance matching can be independently tuned. The leftmost curves are for $V_f = -17$ V while the rightmost curves are for $V_f = 35$ V which changes the resonance frequency by about 15 MHz. While for $V_m = 0$ V the circuit is not perfectly impedance matched (red curves), this can be achieved for an appropriate choice of $V_m$ (black curves). \textbf{(d)} Phase response as a function of magnetic field with the varactors tuned to impedance matching. No change is observed in the resonance frequency or matching over a range of 9 T. The curves are offset for clarity.}
\end{figure}

\textbf{Charge sensitivity} To demonstrate the relevance of impedance matching using SrTiO$_3$ varactors for improving rf readout, we measure the charge sensitivity of a carbon nanotube quantum dot at $T=6$ mK. For these measurements we set the back gate voltage to the steepest slope of a conductance peak (position 'B' in the graph of Fig.~1a) and modulate the back gate voltage with a small sinusoidal voltage $V_{rms}=10$ $\mu$V and modulation frequency $f_m=520$ Hz. This gate modulation and corresponding changes in the nanotube resistance cause an amplitude modulation of the carrier signal. In the power spectrum of the reflected signal shown in Fig.~4a this results in sidebands at $f_c \pm f_m$. The signal-to-noise ratio (SNR) is then determined from the height of the sidebands with respect to the noise floor. The expected SNR is given by the following relation:
\begin{equation}\label{SNR}
\textrm{SNR}=\bigg|\frac{\partial \Gamma}{\partial R}\Delta R \bigg|^2 \frac{P_C}{P_N}
\end{equation}
where $\Delta R$ is the device resistance modulation and $P_C$ and $P_N$ are the applied rf carrier power at the detection circuit and the noise power added to the signal, respectively \cite{Muller2013}. In our measurement set-up the noise is dominated by the low-temperature amplifier in the amplification chain and can be expressed as $P_N = k_B T_N \Delta f$, where $\Delta f$ is the measurement resolution bandwidth. The varactors allow us to improve the SNR by tuning the device towards impedance matching where the system is most sensitive to changes in the device resistance. In other words, the varactors allow us to optimize $|\partial \Gamma / \partial R|$. Detailed simulations of the SNR for our circuit are provided in Supplementary Section II.

\begin{figure*}
\includegraphics[width=142mm]{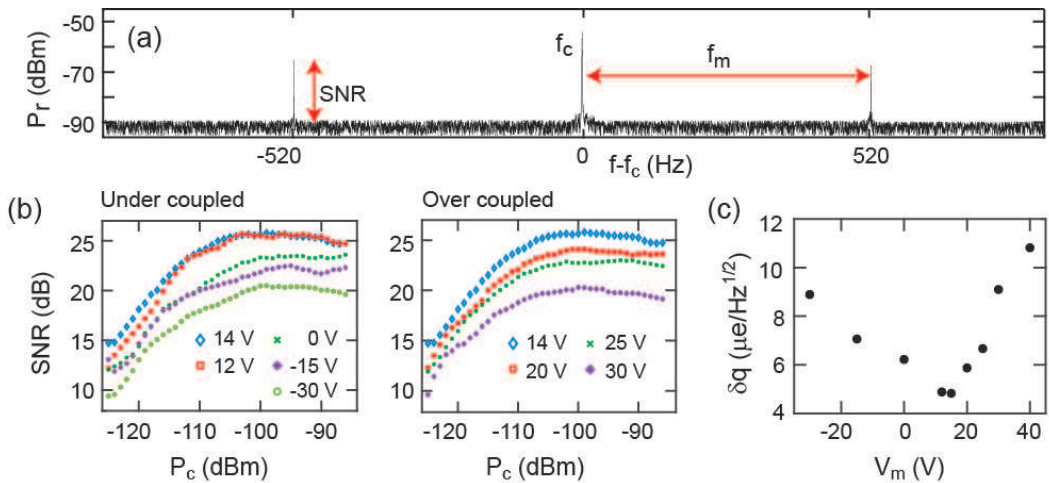}
\caption{\label{Fig4} \textbf{(a)} Reflected power spectrum close to matching showing the carrier frequency $f_c$ and sidebands resulting from a back gate voltage modulation with frequency $f_m=520$ Hz. The resolution bandwidth $\Delta f = 3.4$ Hz. The signal-to-noise ratio (SNR) is obtained from the height of the sidebands with respect to the noise background as indicated. \textbf{(b)} SNR as a function of carrier power for different settings of the voltage $V_m$ on the matching varactor. The voltage on the varactor used for frequency tuning $V_f$ is set to zero. The curves in the left (right) hand panel correspond to under (over) coupling to the $50$ $\Omega$ transmission line. Impedance matching is observed here for $V_m \approx 14$ V. \textbf{(c)} Corresponding charge sensitivities as a function of $V_m$. }
\end{figure*}

The improvement in SNR by tuning the device towards impedance matching is illustrated in Fig.~4b which shows the measured SNR as a function of applied power carrier for different settings of the voltage $V_m$ on the matching varactor. For all $V_m$ curves, the SNR first increases with increasing carrier power consistent with Eq.~1. For very large applied powers the SNR decreases again due to the onset of non-linearity of the device resistance and power broadening of the electron transition by the rf carrier \cite{Ahmed2018}. As expected, the best results are obtained when the device is tuned towards impedance matching. This is most clearly shown when we plot in Fig.~4c the results for the charge sensitivity $\delta q$ which is related to the SNR using the following relation \cite{Brenning2006}:
\begin{equation}\label{sensitivty}
\delta q= \frac{\Delta q_{rms}}{10^{\textrm{SNR}/20}\sqrt{2\Delta f}}
\end{equation}
where $\Delta f = 3.4$ Hz is the resolution bandwidth used here and $\Delta q_{rms}$ is the gate charge induced by the oscillating voltage on the back gate. The latter is obtained from $V_{rms}$ and the known gate capacitance $C_g = e/\Delta V_g$, where $e$ is the electron charge and $\Delta V_g$ the gate voltage difference between two Coulomb blockade conductance peaks as shown in the graph of Fig.~1a. For our carbon nanotube quantum dot this yields $\Delta q_{rms}=2.4\times10^{-4} e$.

Each point in Fig.~4c corresponds to the charge sensitivity obtained for a different varactor voltage $V_m$ and a carrier power $P_c = -99$ dBm which is the approximate power at which the SNR peaks in each of the plots of Fig.~4b. We observe a minimum charge sensitivity $\delta q \sim 4.8$ $\mu e/\textrm{Hz}^{1/2}$ when the device is tuned towards impedance matching at $V_m \sim 14$ V, which is among the best reported to date. As follows from Eq.~1, this can be further improved by using an amplifier with a lower noise temperature such as a Josephson parametric amplifier \cite{Simbierowicz2018}. We note that the SrTiO$_3$ varactors can also be used to optimize quantum \textit{capacitance} ($C_q$) readout such as required for, e.g., quantum dot charge or spin qubit devices. An example of rf capacitance measurements of a carbon nanotube double quantum dot in which the varactors are used, as well as a detailed comparison of the data with model calculations of the rf readout, are provided in Supplementary Section III. Following a similar analysis as presented above for charge sensitivity, we are able to demonstrate an excellent capacitance sensitivity of 0.04 aF/Hz$^{1/2}$ using the varactors.

\section{Discussion}

The work described here shows that SrTiO$_3$ varactors work effectively at mK temperatures. This is important as it allows impedance matching and resonator frequency tuning at the temperature range that is most relevant for solid-state quantum devices. Impedance matching significantly improves readout sensitivity, as demonstrated here by using the varactors in a matching circuit coupled to a carbon nanotube quantum dot charge detector. The ability to tune resonator frequencies is important for scalable multiplexed quantum architectures, reducing the amount of connections and wiring needed. Moreover, our work is readily extended to other capacitance or frequency ranges. For example, while the parallel plate varactor geometry used in this work shows capacitance tunability between about 50 pF to 15 pF, it would be straightforward to customize the design for different capacitance values by using different area sizes or interdigitated electrodes. We furthermore expect the varactors to be able to operate well into the GHz frequency range, ultimately limited by the soft optical mode phonon frequency \cite{Worlock1967}. Additional advantages of the SrTiO$_3$ varactors are that they are compact and easily fabricated, do not add significant complexity to an experimental set-up, and are magnetic-field insensitive.

Further improvements can be made by reducing the varactor losses in the detection circuit, for example by using other quantum paraelectrics \cite{Krupka1994,Tagantsev2003} such as potassium tantalate (KTaO$_3$). While SrTiO$_3$ was chosen in this work for the large tunability of its relative permittivity, KTaO$_3$ has a loss tangent that is predicted to be two orders of magnitude smaller than that of SrTiO$_3$ at temperatures below 1 K, at the expense of a somewhat lower electric-field tunability \cite{Geyer2005}. The choice of material then depends on the relative importance (large tunability or low loss tangent) for the specific application.

\section{Methods}

\textbf{Carbon nanotube device fabrication}. Single-wall carbon nanotubes were grown by chemical vapor deposition on degenerately doped SiO$_2$ substrates terminated with a 280 nm dry thermal oxide. The room temperature resistivity of the Si wafers is $\rho <1.5$ m$\Omega$ cm. The carbon nanotubes had a natural abundance carbon isotope ratio (98.89\% $^{12}$C), and were distributed at a concentration of approximately one nanotube per 20 ${\mu m}^2$ on the substrate. Device fabrication consisted of two electron-beam lithography and metal evaporation stages using 5\% (by weight) of 495K PMMA (polymethyl methacrylate) dissolved in anisole as a resist. During the first stage, alignment marks and bond pads were fabricated on the substrates using 5/60 nm of Ti/Au. The carbon nanotubes were subsequently located with respect to the alignment marks using atomic force microscopy. During the second lithography stage, the source and drain electrodes were defined using 5/60 nm of Ti/Au. A wedge bonder is used to connect the various bond pads of the device to the sample holder using Au wires of 25 $\mu$m in diameter.\\

\textbf{SrTiO$_3$ varactor fabrication}. The back side of a 3$\times$3 mm strontium titanate substrate (0.5 mm thickness, single-crystal (001), single-side polished, TiO$_2$-terminated, purchased from SurfaceNet) is metallised with 5/60 nm of Ti/Au. The top (polished) side of the substrate is coated with a double layer of photoresist: LOR10B ($\sim$1 $\mu$m thickness, baked at 190 $^{\circ}$C for 10 minutes), and S1805 ($\sim$0.5 $\mu$m thickness, baked at 115 $^{\circ}$C for 1 minute). A quartz-chrome photomask is used to expose square pads on the photoresist of area $120\times120$ $\mu$m. The substrate is developed in MF-26A for 45 seconds. The square pads are metallised with 5/60 nm of Ti/Au. The SrTiO$_3$ is then sonicated in 1165 Remover in a heat bath at 80 $^{\circ}$C for lift off.

\medskip

Both varactors used in the measurements are fabricated on the same SrTiO$_3$ substrate and are separated by about 2 mm. The back side of the SrTiO$_3$ substrate is directly connected to a printed-circuit board (PCB) using silver paste and annealed for 5 minutes at 120 $^{\circ}$C to ensure good conductance. The square pads on the top side are bonded to the PCB using Au wires of 25 $\mu$m in diameter. The PCB is enclosed in a brass sample holder connected to a cold finger at the mixing chamber of a dilution refrigerator. Sub-miniature push-on (SMP) connectors on the PCB are used to connect to the coaxial measurement lines.\\

\textbf{Experimental methods}. Experiments were carried out in a dry dilution refrigerator with a base temperature of 6 mK. The measurement circuit included several dc lines which were thermally anchored and extensively filtered at various temperature stages yielding a carbon nanotube electron temperature $\sim 12$ mK. The radio-frequency detection circuit was connected to the device as shown in the schematics of Fig.~1a. Briefly, an attenuated radio-frequency signal is directed to the device source electrode through the coupled port of a directional coupler. Using the directional coupler, the reflected signal is extracted. The signal is first amplified by a cryogenic amplifier anchored at 3 K and measured at room temperature with a vector network analyser. The reflected power spectrum (Fig.~4) was measured with a high-frequency lock-in amplifier (Zurich Instruments UHFLI). The noise temperature of the cryogenic amplifier is specified as $T_N \sim 5$ K at 200 MHz. Bias tees allowed for both rf and dc signals to be applied to the device source electrode and SrTiO$_3$ varactors. The three bias tees consisted of a resistor, capacitor and inductor, of values $R_B$ = 1 k$\Omega$, $C_B$ = 100 pF and $L_B$ = 470 nH.\\
\\
\noindent For the model fits in Fig.~2, an expression for the reflected coefficient was obtained in terms of the capacitance and effective resistances of the varactors. The expression was used to fit the data using the Levenberg-Marquardt algorithm. The model is approximated linearly and with each successive iteration the unknown parameters are refined to give an ideal fit.\\
\\
\emph{Acknowledgments--} We thank Christian Sch\"{o}nenberger for access to carbon nanotube growth facilities, and Pavlo Zubko and Marios Hadjimichael for providing the SrTiO$_3$ crystals and help with photolithography, respectively. We gratefully acknowledge funding from the European Research Council (ERC), Consolidator grant agreement no. 648229 (CNT-QUBIT). A.B. is thankful for financial support by the Swiss Nanoscience Institute (SNI) and the Swiss National Science Foundation.

% End of Main text
%---------------------------------------------------------------------------

\onecolumngrid
\newpage

\setcounter{page}{1}
\thispagestyle{empty}

\begin{doublespace}

\begin{center}
\textbf{{\large Supplementary Information for\\
"Quantum paraelectric varactors for radio-frequency measurements at mK temperatures"}}\\
\bigskip
P. Apostolidis,$^{1,2}$ B. J. Villis,$^{1,2}$ J. F. Chittock-Wood,$^{2}$ A. Baumgartner,$^{3}$ V. Vesterinen,$^{4}$ S. Simbierowicz,$^{4}$ J. Hassel,$^{4}$ and M. R. Buitelaar$^{1,2}$\\

\textit{$^{1}$London Centre for Nanotechnology, University College London, London WC1H 0AH, United Kingdom}

\textit{$^{2}$Department of Physics and Astronomy, University College London, London WC1E 6BT, United Kingdom}

\textit{$^{3}$Department of Physics, University of Basel, Klingelbergstrasse 82, 4056 Basel, Switzerland}

\textit{$^{4}$VTT Technical Research Centre of Finland, P.O. Box 1000, FI-02044 VTT, Finland}

\end{center}

\setcounter{figure}{0}

\renewcommand{\figurename}{Figure S}

\vspace{0cm}

\section{I Matching Network Simulation}

\vspace{0.2cm}

\noindent The circuit model used to simulate the data is shown in Supplementary Figure \ref{Circuit}. The two SrTiO$_3$ varactors are modelled as lumped elements with capacitance values $C_m$ and $C_f$ which are tuned by the control voltages $V_m$ and $V_f$, respectively. Dissipative losses are modelled by the resistors $R_m$ and $R_f$. The control voltages are applied via bias tees that consist of a 1 k$\Omega$ resistor, a 100 pF coupling capacitor and 470 nH inductor. A further bias tee is included that allows the application of a source-drain bias voltage $V_{sd}$ over the device. The device is modelled by a resistor $R_d$ in parallel with a capacitance $C_p$, dominated by parasitics, and connected to the circuit via a 320 nH inductor. With the carbon nanotube device tuned in the Coulomb blockade regime (position '$A$' in Fig.~1 in the main text), we assume $R_d$= 1 G$\Omega$. The capacitance $C_p$ was measured independently at 6 mK during a different cool-down, for which the varactors were bypassed, yielding $C_p$=3.2 pF.\\
\\
\noindent To take into account the self capacitance of the inductors, these were modelled to be parallel with a capacitance and resistor as indicated by the rightmost panel in Supplementary Figure \ref{Circuit}.  The values for $C_L$ and $R_C$ were taken from the inductor datasheets (model: Coilcraft 0805CS-331): $L$=320 nH, $R_c$=31 $\Omega$ and $C_L$=0.096 pF and (model: Coilcraft 0805CS-471): $L$=470 nH, $R_c$=66 $\Omega$ and $C_L$=0.132 pF. The datasheets also include values for frequency-dependent series resistors that model wire losses (of order 6-8 $\Omega$ at a frequency of 175 MHz). We were unable to fit the measured data with these values and suggest that the datasheet - obtained at room temperature - overestimates the values of the series resistors when measured at 6 mK. Instead we modelled the dissipative losses of the 320 nH inductor by incorporating this in the fit parameter $R_f$. For the 470 nH bias tee inductors these values are not significant for our model as the inductors are in series with 1 k$\Omega$ resistors.\\
\\
\noindent Using the model we calculate the reflection coefficient $\Gamma$ given by \cite{Pozar}:
\begin{equation}
\label{eq:Gamma}
\Gamma(f)=\frac{Z(f)-Z_0}{Z(f)+Z_0}
\end{equation}

\begin{figure}[t]
	\begin{center}
		\includegraphics[width=145mm]{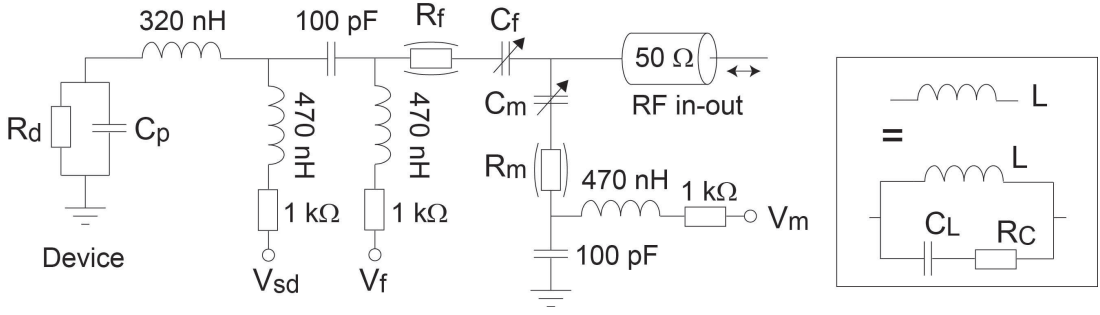}
\vspace{0.3cm}
		\caption {\label{Circuit} Circuit model used to simulate the matching network and device. The varactors are taken as lumped elements with voltage tunable capacitances $C_f, C_m$ and effective series resistances $R_f, R_m$, respectively. The nanotube quantum dot device is parameterised by a resistance $R_d$ and parasitic capacitance $C_p$. To take into account the self capacitance of the inductors, these were modelled to be parallel with a capacitance and resistor as indicated by the rightmost panel.}
	\end{center}
\end{figure}

\begin{wrapfigure}[17]{r}{195pt}
	\begin{center}
	\includegraphics[width=59mm]{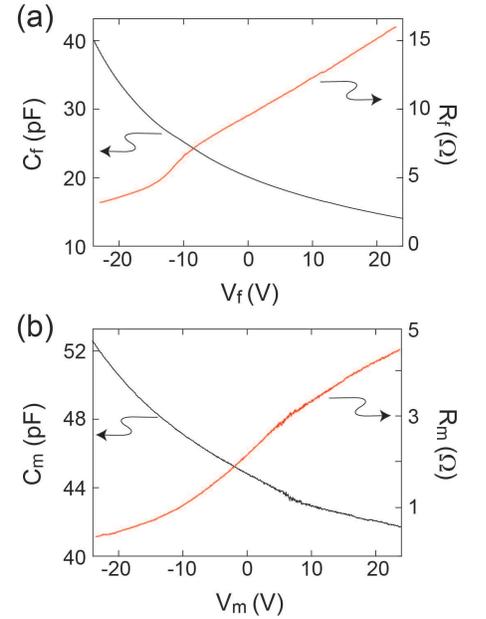}
\end{center}
\vspace{-0.3cm}
	\caption {\label{CandR} \textbf{(a)} Varactor capacitance $C_f$ and effective series resistance $R_f$ as a function of tuning voltage. \textbf{(b)} Idem as panel (a) for $C_m$ and $R_m$.}
\end{wrapfigure}

\noindent where $Z(f)$ is the total impedance of the network in Supplementary Figure \ref{Circuit} for a given frequency $f$ and $Z_0$=50 $\Omega$ is the line impedance to which it is connected. The reflection coefficient is measured using a vector network analyzer where we take into account a constant insertion loss resulting from attenuation in the measurement lines and amplifier gains. In our measurement set-up, the total attenuation in the measurement lines amounted to 85 $\pm$ 4 dB: 40 dB from fixed attenuators at room temperature and 45 dB in our dilution refrigerator which includes fixed attenuators at various temperature stages, a directional coupler and losses in the coaxial lines. Amplification in the return line is provided by a cryogenic amplifier operated at 3 K, providing 45 dB gain, and a further amplification stage at room temperature providing 60 dB gain. For a perfectly matched network the reflection coefficient is zero. In practise the minimum detectable signal amplitude is limited by the noise floor of the network analyzer.

\noindent The data fits in Fig.~2a in the main text were obtained from the expression for $\Gamma$ above using $C_f$ and $R_f$ as fit parameters for all frequency sweeps, as $V_f$ was varied. Other parameters, including an overall offset to account for the insertion loss, were kept fixed. We used the Levenberg-Marquardt algorithm  for the fits (also known as the damped least-squares method), that analyses a set of $m$ observations with $n$ unknown parameters where $m>n$ \cite{Marquardt} from the SciPy library (open-source scientific software environment for Python). The model is approximated linearly and with each successive iteration, the unknown parameters are refined to give an ideal fit. Relevant fitting constraints were introduced to ensure the fitting parameters made physical sense. The results for $C_f$ and $R_f$ as a function of the corresponding voltage bias $V_f$ are shown in Supplementary Figure \ref{CandR}a (identical to Fig.~3a in the main text).
\vspace{0.3cm}

\noindent The data fits for $C_m$ and $R_m$, corresponding to Fig.~2b in the main text, followed the same procedure but we note that the best results were obtained assuming a (small) linear cross-talk between the two varactors: 0.13 pF/V increase in $C_f$ and 0.07 $\Omega$/V decrease in $R_f$ for $V_m =-24$ to 24 V. Crosstalk might be the result of the proximity of the two varactors on the same SrTiO$_3$ substrate - although this could not be confirmed using a finite-element (COMSOL) simulation of the device which predicted negligible crosstalk. Simulations for $C_f$ and $R_f$ are therefore taken as more reliable when compared to $C_m$ and $R_m$. The results for $C_m$ and $R_m$ as a function of the corresponding voltage bias $V_m$ are shown in Supplementary Figure \ref{CandR}b.

\section{II Charge Sensitivity: $|\Gamma|, |\Delta \Gamma|$, and SNR simulations}

\noindent The ability to tune the matching capacitance of the the rf reflectometry readout setup allows optimisation of the measured signal-to-noise ratios (SNR) and therefore of the achievable charge sensitivity. This can be understood by considering the expected SNR:
\begin{equation}
\label{eq:SNR}
\textrm{SNR}=\bigg|\frac{\partial \Gamma}{\partial R}\Delta R \bigg|^2 \frac{P_C}{P_N}
\end{equation}
\noindent where $R$ is the device resistance and $P_C$ and $P_N$ are the applied rf carrier power at the input of the tuning circuit and the noise power added to the signal, respectively \cite{Muller}. In our set-up, the noise is dominated by the low-temperature amplifier (CITLF1) mounted at the 3K stage of the dilution refrigerator which has a noise temperature $T_N = 5$ K. It follows from Supplementary Eq. 2 that there are several ways to improved the SNR:

\begin{figure}
	\centering
	\includegraphics[width=175mm]{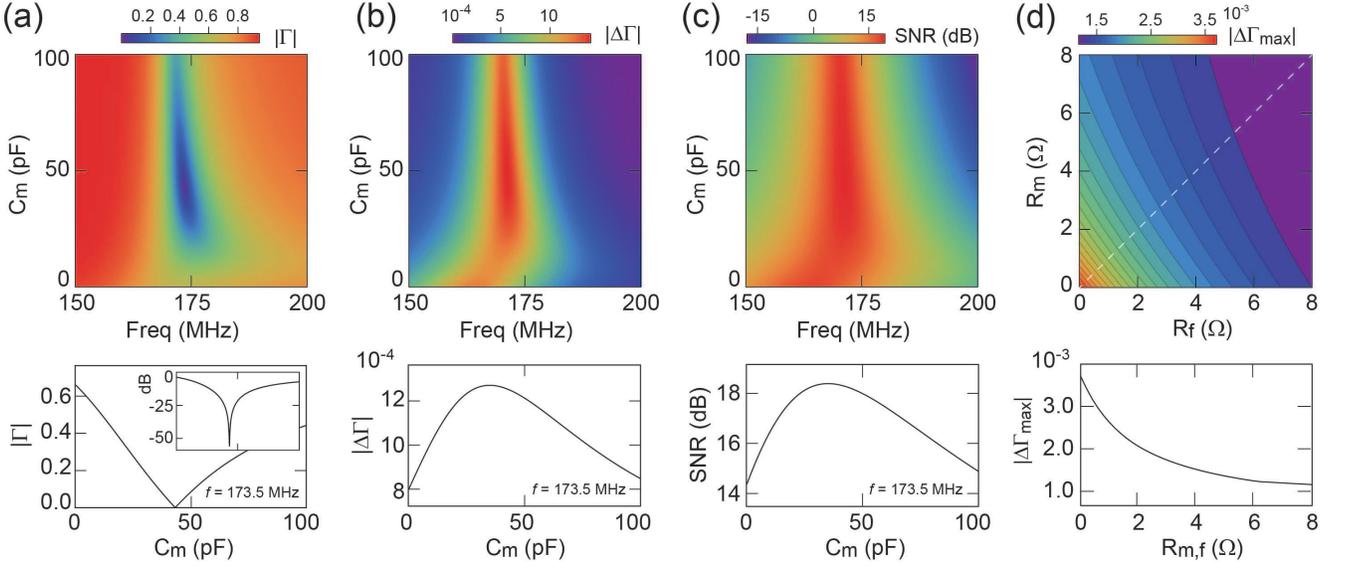}
	\caption {\label{QD} \textbf{Charge sensitivity simulations} \textbf{(a)} Reflectance magnitude $|\Gamma|$ as a function of frequency and matching capacitance $C_m$ using the circuit parameters described in the text. The line graph shows the corresponding trace at a frequency $f=173.5$ MHz (inset: same graph on dB scale)  \textbf{(b)} Change in reflectance magnitude $|\Delta \Gamma|$ as a function of frequency and matching capacitance $C_m$. \textbf{(c)} SNR as a function of frequency and matching capacitance $C_m$. \textbf{(d)} Contour plot showing the maximum $|\Delta \Gamma_{max}|$ observed in the $C_m$-freq plots as in panel (b), calculated for a range of $R_m$ annd $R_f$ between 0 and 8$\Omega$. The line scan shows $|\Delta \Gamma_{max}|$ along the white dashed line in the contour plot.}
\end{figure}

\begin{itemize}%[rightmargin=\dimexpr\linewidth-16.5cm-\leftmargin\relax]
  \item \textbf{$P_C$}: The SNR is directly proportional to the applied power as also observed in the range $-125$ to $-110$ dBm in Fig.~4b in the main text. In practise, however, the power that can be applied is limited by the non-linearity of the quantum dot conductance (or any other sensor) at large voltage bias and by thermal broadening. This yields an optimal SNR, seen around $P_C = -99$ dBm in Fig.~4b in the main text.
  \item \textbf{$P_N$}: Reducing the noise power $P_N=k_B T_N \Delta f$ improves the SNR. For a given resolution bandwidth $\Delta f$, this is achieved by using an amplifier with a lower noise temperature [see Supplementary Section III below].
  \item \textbf{$|\partial \Gamma / \partial R|$}: Using the varactors to optimize the sensitivity of the reflection coefficient with respect to changes in the device resistance improves the SNR. This is illustrated in Supplementary Figures 3a-c showing the dependence of $|\Gamma|, |\Delta \Gamma|$, and SNR as a function of frequency and $C_m$ for the circuit used in the experiment [Supplementary Figure \ref{Circuit}]. Here,
      \begin{equation}
      |\Delta \Gamma|=|\frac{\partial \Gamma}{\partial R}\Delta R|
      \end{equation}
      where we use $\Delta R = 500$ k$\Omega$, which is the change in device resistance resulting from the gate modulation $V_{rms}$. For the calculation of the SNR we use an applied power $P_C = -110$ dBm, $T_N =5$ K and $\Delta f=3.4$ Hz. We furthermore assume fixed dissipative losses $R_m=3 \Omega$ and $R_f=3 \Omega$. The simulations correspond well with the measured data in Fig.~4 main text, showing a clear improvement in the SNR as the circuit is tuned towards matching \footnote{Note that an applied power of -110 dBm is at the upper limit in the data of Fig.~4b in the main text where the linear relation between SNR and $P_C$, as expressed in Supplementary Eq. 2, still holds. The maximum SNR seen in the data at this power is 23 dB; while for the model calculations this is 19 dB - that is, about 4 dB less; just within the uncertainty estimate of the calculations.}.
\end{itemize}

\noindent For simplicity the dissipative varactor losses were kept constant in the simulations of Supplementary Figures 3a-c. To illustrate to which extent the maximum achievable sensitivity depends on $R_m$ and $R_f$ we show in Supplementary Figure 3d the maximum $|\Delta \Gamma_{max}|$ that is observed in plots of $C_m$ versus frequency (such as Supplementary Figure 3b) for a range of $R_f$ and $R_m$ between 0-8 $\Omega$. The figure shows that $|\Delta \Gamma_{max}|$ monotonically increases as the losses decrease, see line graph. The improvement is modest for the upper half of the range but is of order 3-4 for the lower half, yielding a potential improvement in the SNR of order 10 dB given that the SNR $\propto |\Delta \Gamma|^2$. Reducing the varactors losses, e.g. by using the quantum paraelectric KTaO$_3$ which is expected to have a lower loss tangent by two orders of magnitude at mK temperatures \cite{Geyer}, would therefore be an interesting option to explore.

\section{III Capacitance Sensitivity: $|\Gamma|, |\Delta \Gamma|$, and SNR simulations}

\begin{figure}
	\centering
	\includegraphics[width=120mm]{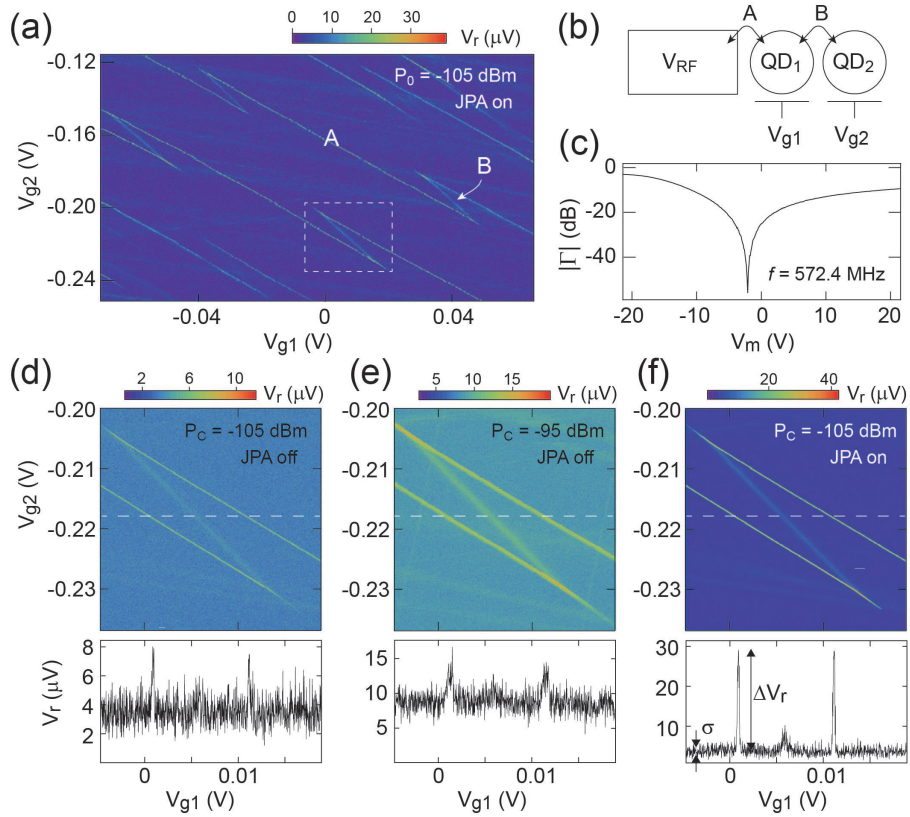}
	\caption {\label{JPA} \textbf{(a)} Double quantum dot charge stability diagram showing reflectance magnitude $|\Gamma|$ as a function of plunger gate voltages $V_{g1}$ and $V_{g2}$. Labels $A$ and $B$ indicate charge transitions between the lead and left quantum dot and between the left quantum dot and right quantum dot, respectively \textbf{(b)} Schematic of the double quantum dot device. \textbf{(c)} Reflectance magnitude $|\Gamma|$ as a function of matching varactor voltage $V_m$ for a frequency $f_c = 572.4$ MHz. Impedance matching is observed around $V_m = -2.1$ V. \textbf{(d)} Part of the double quantum dot stability diagram as indicated by the white dashed lines in panel (a) showing reflectance magnitude $|\Gamma|$ as a function of the plunger gate voltages $V_{g1}$ and $V_{g2}$ for various setting of the carrier power $P_C$ and JPA settings (on/off) as indicated in the insets. The line traces show $|\Gamma|$ as measured along the white dashed lines in the figures.}
\end{figure}

\noindent The varactors also allow optimization of rf-reflectometry readout of double quantum dot devices - of interest as charge or singlet-triplet spin qubits. To demonstrate this, we show in Supplementary Figure \ref{JPA} measurements of a carbon nanotube double quantum dot coupled to a single electrode and controlled by two separate gate electrodes. For a double quantum dot geometry, charge transitions - e.g. the ability of an electron to move between the left to right quantum dot - in response to an applied voltage, correspond to a change in the device reactance (or capacitance) rather than its resistance\footnote{This assumes that tunnel rates are larger than the rf drive frequency which is the case here; see also e.g. Ref.~\cite{Chorley}.}. This capacitance change is measured using the reflectometry set-up described in our manuscript which includes the varactors. Similarly to Supplementary Eq. 2 above, the SNR is then given by:
\begin{equation}
\label{eq:SNR}
\textrm{SNR}=\bigg|\frac{\partial \Gamma}{\partial C}\Delta C \bigg|^2 \frac{P_C}{P_N}
\end{equation}
As shown below, the varactors allow us to optimize the SNR and readout of double quantum dot charge states on state-of-the-art $\mu$s timescales which is significantly faster than typical spin qubit relaxation times.\\
\\
\noindent For the measurements in Supplementary Figure \ref{JPA} we used the schematic and set-up as illustrated in Supplementary Figure \ref{Circuit} - and the same varactors as described in the main text. However, as the double quantum dot device has been fabricated on an \textit{undoped} Si/SiO$_2$ substrate, its parasitic capacitance $C_p = 0.27$ pF is significantly lower as compared to the single quantum dot device described in the main text. We furthermore used a resonator inductor with inductance $L=220$ nH, yielding a resonance frequency in the 570-580 MHz range. These values were choosen such that the resonator frequency is compatible with the use of a Josephson parametric amplifier (JPA) additionally incorporated in the set-up \cite{Simbierowicz}. To be able to use the same varactors (as described in the main text) to tune the resonator response through impedance matching in this frequency range we included a fixed 20 pF capacitor in series with $C_m$.\\
%(cite VTT)*************************************
\\
Supplementary Figure \ref{JPA}a shows the charge stability diagram of the double quantum dot as a function of the two plunger gate voltages $V_{g1}$ and $V_{g2}$. The charge transitions that are visible are those between the source electrode and quantum dot one (labelled 'A') and between the two quantum dots (labelled 'B') as also illustrated in the schematic of Supplementary Figure \ref{JPA}b. The resonator response was measured at a frequency of 572.4 MHz and tuned to impedance matching using the varactor voltage $V_m$ as shown in Supplementary Figure \ref{JPA}c. Using the varactors we were able to obtain excellent SNR [29 dB for transition \textit{A} and 14 dB for transition \textit{B}] using an integration time of $\sim 10 \mu$s yielding state-of-the-art capacitance sensitivity. The fast readout is made possible by the varactors which allow SNR optimization in two different ways:

\begin{itemize}%[rightmargin=\dimexpr\linewidth-16.5cm-\leftmargin\relax]
  \item By optimizing $|\partial \Gamma/\partial C|$. The best SNR is obtained around a narrow range of $C_m \sim 18$ pF as also illustrated in the model calculations of Supplementary Figure \ref{DQD}.
  \item By minimizing the background reflectance magnitude $|\Gamma|$ as the device is tuned towards impedance matching. This is important as it allows us to optimize the use of the JPA in the set-up. This type of amplifier is easily saturated (typically around input powers of -120 dBm for a gain of 20 dB) and minimizing $|\Gamma|$ using the varactors allows us to maximize $P_C$ and thus the SNR.
\end{itemize}

\begin{figure}
	\centering
	\includegraphics[width=175mm]{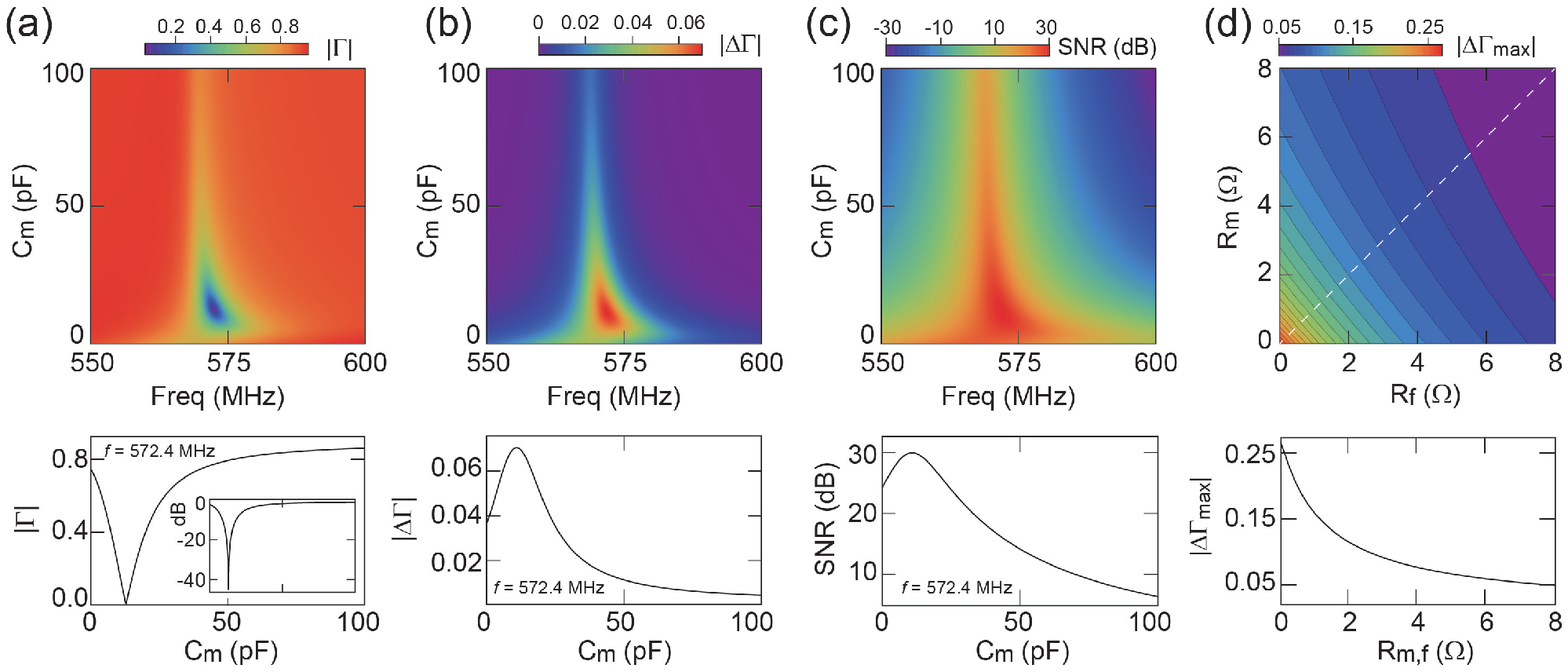}
	\caption {\label{DQD} \textbf{Capacitance sensitivity simulations}: \textbf{(a)} Reflectance magnitude $|\Gamma|$ as a function of frequency and matching capacitance $C_m$ using the circuit parameters described in the text. The line graph shows the corresponding trace at a frequency $f=572.4$ MHz (inset: same graph on dB scale)  \textbf{(b)} Change in reflectance magnitude $|\Delta \Gamma|$ as a function of frequency and matching capacitance $C_m$. \textbf{(c)} SNR as a function of frequency and matching capacitance $C_m$. \textbf{(d)} Contour plot showing the maximum $|\Delta \Gamma_{max}|$ observed in the $C_m$-freq plots as in panel (b), calculated for a range of $R_m$ annd $R_f$ between 0 and 8$\Omega$. The line scan shows $|\Delta \Gamma_{max}|$ along the white dashed line in the contour plot.}
\end{figure}

\noindent The improvement of the SNR using the JPA is further illustrated in Supplementary Figures \ref{JPA}d-f. In each of the plots, the integration time per point is approximately 10 $\mu$s [bandwidth $\Delta f = 78$ kHz]. The leftmost plot shows part of the stability diagram (indicated by the dashed lines in Supplementary Figure \ref{JPA}a) with an input power at the device of -105 dBm when the JPA (which works in reflection) is switched off. Increasing the power $P_C$ (middle) panel does not improve the SNR as it results in (power) broadening of the transition lines. Incorporating the JPA, which has a noise temperature $T_N = 0.1$ K, however, does significantly improve the SNR as seen in the rightmost panel. Expressing the SNR in power, i.e. $(\Delta V_r/\sigma)^2$; see line scans in Supplementary Figures \ref{JPA}d-f, we obtain an SNR of 14 dB for transition $A$ without the JPA (leftmost plot) and an SNR of 29 dB for transition $A$ with the JPA (rightmost plot). The SNR improvement of 15 dB is exactly as expected given the different in noise temperature: $T_N=5$ K without the JPA and $T_N=0.15$ K with the JPA. The latter consists of two components, the noise temperature of 0.1 K of the JPA and the residual contribution of the (now) second-stage amplifier with noise temperature 5 K divided by the 20 dB gain of the JPA.\\
\\
\noindent The obtained SNR is consistent with model calculations of the device and readout circuit as shown in Supplementary Figure \ref{DQD}c. We modeled transition $A$ [electron transfer between the source electrode and left quantum dot] for which we used a capacitance $\Delta C = 0.4$ fF. This value follows from the relation $\Delta C = (e^2 \alpha^2)/(4 k_B T_e)$ \cite{Chorley} where $\alpha=0.2 \pm 0.05$ is a lever arm obtained from the stability diagram that relates voltages on the electrodes to changes in electrochemical potential of the quantum dot and $T_e \approx 50$ mK is the electron temperature of the device in this set up. For the SNR calculations we used an applied power $P_0 = -105$ dBm, bandwidth $\Delta f=78$ kHz and noise temperature $T_N=0.15$ K as used in the experiment. We used the same parameters for the varactors as in similar model calculation for the single quantum dot device described above.\\
\\
\noindent The calculated optimum SNR at impedance matching of 30 dB, see Supplementary Figures \ref{DQD}c, is in good agreement with the experimental value (29 dB) obtained from the data in Supplementary Figures \ref{JPA}f. From this we can calculate the obtained capacitance sensitivity:

\begin{equation}
\label{eq:Capsensitivity}
\delta C= \frac{\Delta C}{10^{\textrm{SNR}/20}\sqrt{2\Delta f}}
\end{equation}

\noindent which for the obtained experimental parameters yields state-of-the-art $\delta C = 0.04$ aF/$\sqrt{\textrm{Hz}}$. Of relevance for quantum readout is also the SNR observed for transition $B$ in Supplementary Figure \ref{JPA}. It is this inter-dot transition that is used for state readout of double quantum dot charge qubits or - in the presence of spin-selective tunneling such as due to Pauli blockade - singlet-triplet spin qubits. For an integration time of 10 $\mu$s we observe an SNR of 14 dB which is already sufficient for single-shot spin qubit readout as typical relaxation times are orders of magnitude larger. Nevertheless, faster readout can be achieved by reducing dissipative losses in the readout circuit.\\
\\
\noindent To illustrate to which extent the maximum achievable sensitivity depends on $R_m$ and $R_f$ we show in Supplementary Figure \ref{DQD}d the maximum $|\Delta \Gamma_{max}|$ that is observed in plots of $C_m$ versus frequency (such as Supplementary Figure \ref{DQD}b) for a range of $R_f$ and $R_m$ between 0-8 $\Omega$. As is the case for the charge sensitivity measurements described above, reducing losses yields a further potential improvement in the SNR of order 10 dB and therefore a corresponding improvement in quantum state readout timescales.

\section{IV Varactor Hysterisis and Resonance Stability}

\noindent Strontium titanate shows hysteretic behaviour \cite{Caviglia,Davidovikj} resulting in a dependence of the varactor capacitance on the voltage sweep history. To illustrate the degree of hysteresis, we plot in Supplementary Figure \ref{ResonanceStability}a for each data point the minimum reflection coefficient magnitude measured over a frequency range of 150 to 200 MHz as a function of $V_m$ and $V_f$, for two different sweep directions, as indicated by the arrows in the top and bottom panels. The measurement in the top panel is taken from the bottom left to the top right corner ($V_m$=-30 V, $V_f$=-24 V to $V_m$=+30 V, $V_f$=+24 V) and vice-versa for the bottom panel. The dark blue line in each plot corresponds to pairs of ($V_f$,$V_m$) where perfect impedance matching is observed - clearly different for the two sweep directions.\\
\\
\noindent Important for device operation is that after tuning (taking into account the sweep history as illustrated above) the varactors are stable over time and do not add noise to the measurements - a possibility given that the varactors are tunable using electric fields and are, therefore, susceptible to charge noise. To illustrate this, we measure at the resonance frequency ($f_0=173.2$ MHz) over a period of one hour with the varactors set to matching conditions, as shown in Supplementary Figure \ref{ResonanceStability}b,c, taking data each minute. In the top panels of Supplementary Figure \ref{ResonanceStability}c we plot the quadratures ($X$,$Y$), using $X+iY = |\Gamma|e^{i\phi}$. The bottom panel of Supplementary Figure \ref{ResonanceStability}c shows the scatter observed in the reflection coefficient, that is, $|\delta \Gamma| = \sqrt{(X-\overline{X})^2+(Y-\overline{Y})^2}$, where $\overline{X}$ and $\overline{Y}$ are the mean values taken over the measurement. No long term drift is observed in $|\delta \Gamma|$.\\
\\
\noindent The measured values for $|\delta \Gamma|$ are consistent with a dominant noise contribution of the low-temperature amplifier (operated at 3 K) in the measurement set-up. This yields an estimate for $|\delta \Gamma|$ of order $\sqrt{P_N/P_C}$, where $P_N = k_B T_N \Delta f$ is the amplifier noise power over a frequency bandwidth $\Delta f$, for a amplifier noise temperature $T_N$, and applied power $P_0$ \cite{Muller}. For the data in Supplementary Figure \ref{ResonanceStability}c we have $T_N \approx 5$ K, a resolution bandwidth $\Delta f = 100$ Hz and applied power $P_C = -95$ dBm, yielding $|\delta \Gamma| \approx 10^{-4}$ consistent with our data. We therefore conclude that the scatter observed in the reflection coefficient is dominated by amplifier noise and other noise sources (including possible noise introduced by the varactors) will have much smaller contributions.

\begin{figure*}
	\centering
	\includegraphics[width=130mm]{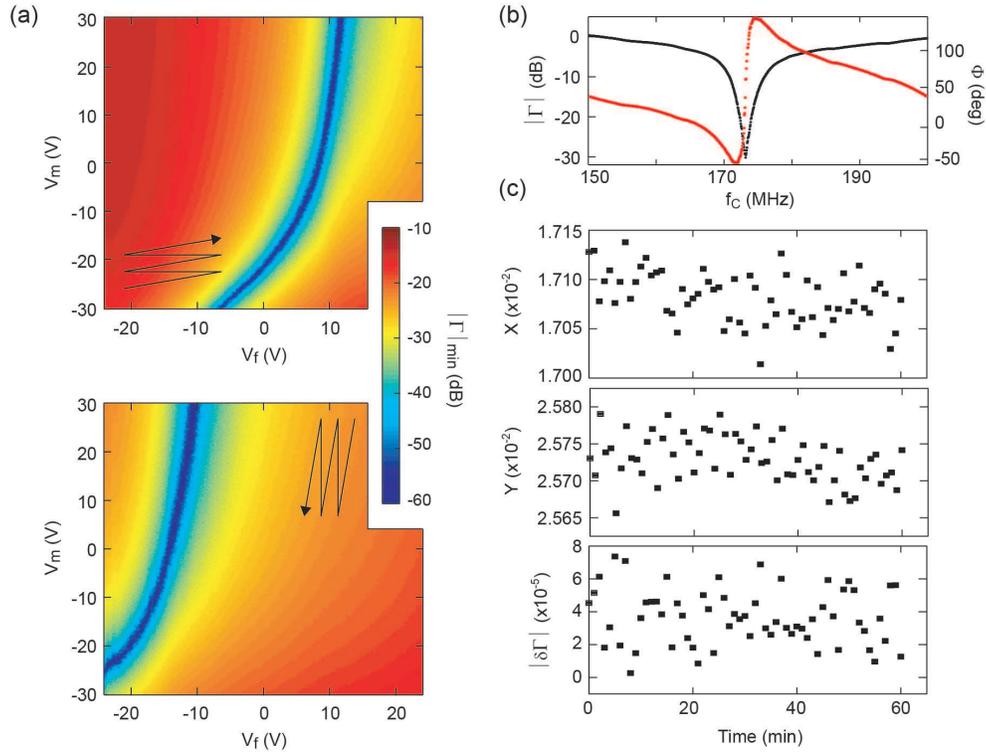}
	\caption {\label{ResonanceStability} \textbf{(a)} Colourscale plots showing, for each data point, the minimum reflectance coefficient magnitude $|\Gamma|$ measured over a frequency range of 150 to 200 MHz as a function of $V_m$ and $V_f$. The top panel shows the result for a sweep direction from the bottom left to the top right corner ($V_m$=-30 V, $V_f$=-24 V to $V_m$=+30 V, $V_f$=+24 V) and vice-versa for the bottom panel. \textbf{(b)} Measured reflection coefficient magnitude and phase response of the device with the varactors set to matching conditions \textbf{(c)} Measured quadratures $X,Y$ (top panels) and variation of the reflection coefficient $\delta \Gamma$ over a one hour time period as described in the text.}
\end{figure*}

\end{doublespace}

%-------------------------------------------------------------

\end{document}